\documentclass[twocolumn]{aastex631}
\usepackage{amsmath}
\usepackage{multirow}
\usepackage{mathtools}
\usepackage[thinc]{esdiff}

\newcommand{\bz}{$\langle B_z \rangle$}

\received{----}
\revised{----}
\accepted{----}
\submitjournal{ApJ}

\begin{document}

\title{Peculiar spectral property of coherent radio emission from a hot magnetic star: \\the case of an extreme oblique rotator}

\shorttitle{Wideband observation of HD 142990}
\shortauthors{Das et al.}

\correspondingauthor{Barnali Das}
\email{Barnali.Das@csiro.au}

\author[0000-0001-8704-1822]{Barnali Das}
\affil{Department of Physics and Astronomy, Bartol Research Institute, University of Delaware, 217 Sharp Lab, Newark, DE 19716, USA}
\affil{CSIRO Space and Astronomy, PO Box 1130, Bentley WA 6102, Australia}

\author[0000-0002-0844-6563]{Poonam Chandra}
\affil{National Radio Astronomy Observatory, 520 Edgemont Road, Charlottesville VA 22903, USA}







\begin{abstract}
We report ultra-wideband (0.4--4.0 GHz) observation of coherent radio emission via electron cyclotron maser emission (ECME) produced by the hot magnetic star HD\,142990. 
With nearly perpendicular rotation and magnetic dipole axes, it represents an extreme case of oblique rotators. The large obliquity is predicted to cause complex distribution of stellar wind plasma in the magnetosphere \citep{townsend2005}. It has been proposed that such a distribution will give rise to non-trivial frequency dependence of ECME \citep{das2020a}.
Indeed we discovered strong frequency dependence of different pulse-properties, such as appearance of secondary pulses, different cut-off frequencies for pulses observed at different rotational phases etc.. But the unique feature that we observed is that while at sub-GHz frequencies, the star appears to produce ECME in the extra-ordinary mode, at GHz frequencies, the mode indicated by the pulse-property is the ordinary mode. By considering the physical condition needed by such a scenario, we conclude that the required transition of the magneto-ionic mode with frequency is unlikely to occur, and the most promising scenario is refraction caused by the complex plasma distribution surrounding the star. 
This suggests that the conventional way to deduce the magneto-ionic mode based on ECME observed at a given frequency is not a reliable method for stars with large misalignment between their rotation and magnetic axes.
We also find that ECME exhibits an upper cut-off at $\lesssim 3.3$ GHz, which is much smaller than the frequency corresponding to the maximum stellar magnetic field strength. 
\end{abstract}

\keywords{masers -- radiation mechanisms: non-thermal -- radio continuum: stars -- stars: early-type -- stars: individual (HD\,142990) -- stars: magnetic field}


\section{Introduction} \label{sec:intro}
Periodic radio pulses produced by electron cyclotron maser emission (ECME) is observed from objects lying at the extreme ends of the stellar spectral classification \citep[e.g.][]{trigilio2000, berger2001}. Towards the top, we have the hot magnetic stars, which are stars of spectral types B and A (no O star was discovered to produce ECME) with large-scale magnetic fields, constituting 10\% of the AB star population \citep[e.g.][]{grunhut2017}. Towards the bottom, we have ultracool dwarfs and brown dwarfs. For the latter, the discovery of such emission came as a surprise \citep[e.g.][]{berger2001,hallinan2006} since it was difficult to understand the generation of magnetic fields in fully convective objects,
and the source of energetic particles were also unknown \citep[e.g.][]{williams2018}. On the other hand, for hot magnetic stars, the basic ingredients for ECME: magnetic field and energetic particles, are easily provided (thanks to their strong stellar wind), thus suggesting that many hot magnetic stars will be able to produce the emission. Despite that, within two decades, only 7 such stars were discovered: CU\,Vir \citep{trigilio2000}, HD\,133880 \citep{chandra2015,das2018}, HD\,142301 \citep{leto2019}, HD\,142990 \citep{lenc2018,das2019a}, HD\,35298 \citep{das2019b}, HD\,147933 \citep{leto2020} and HD\,147932 \citep{leto2020b}. This mystery was resolved only recently when, with the help of a systematic survey at sub-GHz frequencies with the upgraded Giant Metrewave Radio Telescope (uGMRT), \citet{das2022} 
inferred that at least 35\% of the hot magnetic stars are capable of producing ECME. Thus the phenomenon is probably ubiquitous among the population of hot magnetic stars, consistent with expectation. In 2021, \citeauthor{das2021} introduced the term `Main-sequence Radio Pulse emitter' (MRP) to describe hot magnetic stars that emit ECME.

Although a key mystery regarding MRPs was resolved, most of the characteristics of the phenomenon are still poorly understood. For example, what stellar parameters are important to determine the ECME luminosity is not yet clear. By comparing spectral ECME luminosity with that of incoherent non-thermal radio luminosity at 700 MHz, \citet{das2022c} inferred that for late B and A type stars, the same set of physical parameters \citep[magnetic field, stellar radius, rotation period;][]{leto2021,shultz2022,owocki2022} appear to drive both incoherent and coherent radio emission; however, the correlation breaks for early-B type stars and the reason behind is unknown. A more meaningful comparison can be performed by increasing the sample size and by comparing luminosities rather than spectral luminosities. But for that we need the information about ECME spectra. 
So far, such wideband (from sub-GHz to GHz frequencies) observations have been reported for only three hot magnetic stars: HD\,133880 \citep{das2020b}, CU\,Vir \citep{das2021} and HD\,35298 \citep{das2022b}. These observations, however, introduced more unexplained characteristics of the phenomenon. For example, in all cases the upper cut-off frequencies were found to be much lower than that corresponding to their respective maximum surface magnetic field strengths. Also, the cut-offs were found to be different for emission arising from the same magnetic hemisphere but observed at different rotational phases \citep[e.g.][]{das2020b}. In addition, observation of CU\,Vir for its full rotation cycle revealed that not just the pulse-properties, but the number of pulses observable per cycle is also a function of frequency \citep{das2021}. None of these features were consistent with the originally conceived notions about spectral properties of ECME from hot magnetic stars \citep{trigilio2011,lo2012,leto2016}. It was later proposed that such anomalous characteristics are signatures of stellar wind plasma distributed in an azimuthally asymmetric fashion in the magnetosphere \citep{das2020a}.

The discovery that spectral properties of MRPs are more complex than what had been thought adds further motivation to observe more MRPs over wide frequency ranges. The MRPs with large misalignment between magnetic and rotation axes are especially of interest, since these are the stars where the stellar wind plasma is expected to have complex distribution \citep{townsend2005}. In this paper, we report observation of ECME over 0.4--4.0 GHz for the MRP HD\,142990, in which the magnetic dipole axis and the rotation axis are inclined by an angle $\approx 90^\circ$ \citep{shultz2019c}. ECME from the star was first speculated by \citet{lenc2018} at 200 MHz, which was then confirmed by \citet{das2019a} at 600--700 MHz. Note that \citet{das2019a} already noted that the star's ECME pulse-profiles are peculiar as compared to that observed for other MRPs in the similar frequency range.


This paper is structured as follows: in the next section we describe our observation campaign and data analysis (\S\ref{sec:observation}), this is followed by the results (\S\ref{sec:results}), and discussion (\S\ref{sec:discussion}). We present our conclusion in section \S\ref{sec:conclusion}.

\begin{figure}
    \centering
    \includegraphics[width=0.45\textwidth]{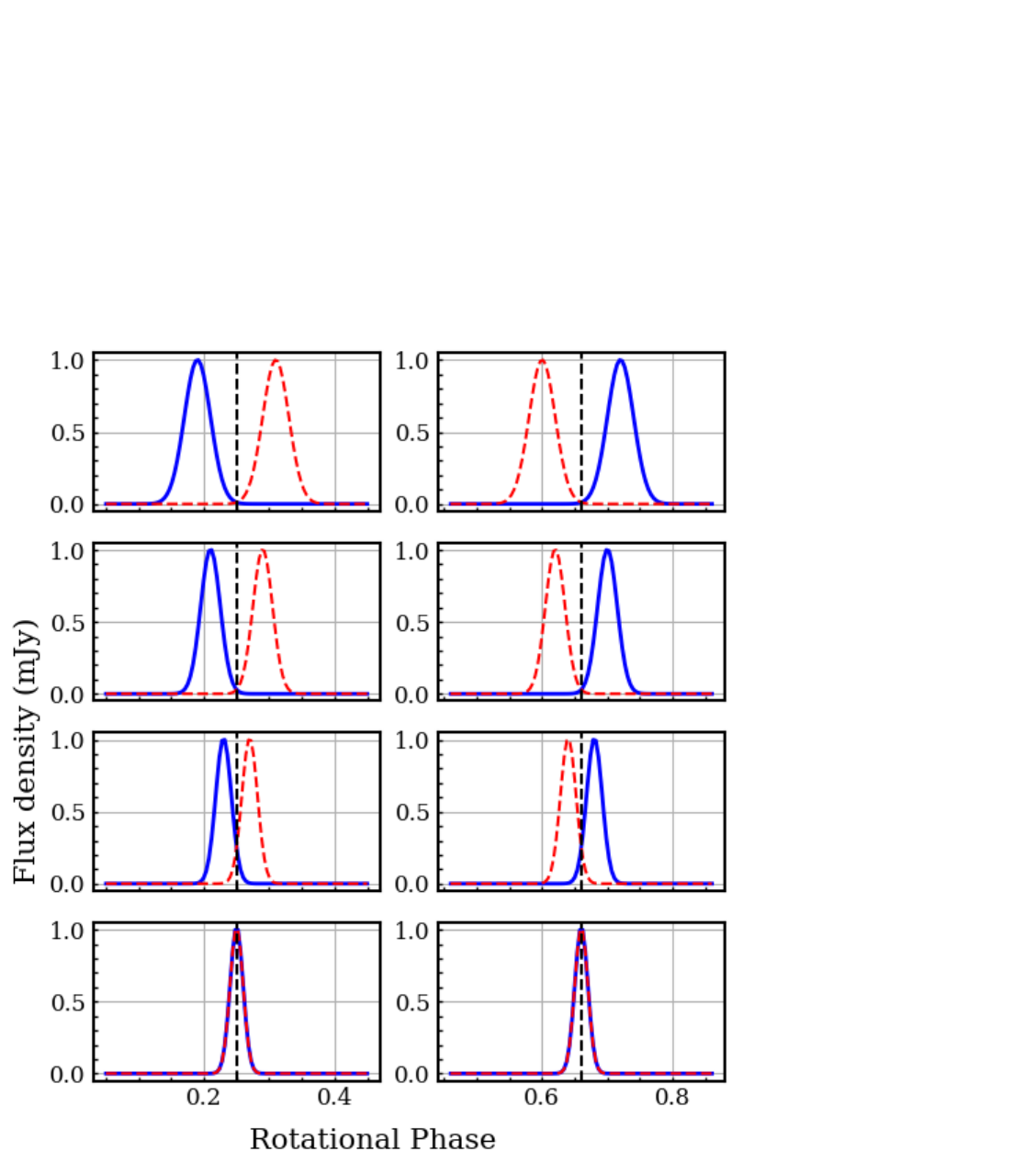}
    \caption{Cartoon diagram illustrating the frequency dependence of ECME lightcurves as a function of observing frequencies (frequency increases from top to bottom) under the scenario proposed by \citet{trigilio2011}. The red and blue colors represent ECME originating at northern and southern magnetic hemispheres respectively, and the vertical dashed lines indicate the respective magnetic nulls. The left panels correspond to null 1, where \bz~changes its sign from negative to positive, and the right panels correspond to null 2, where \bz~changes its sign from positiv eto negative. Note that the pulse-widths are arbitrarily assigned, but motivated from observation that the pulses at lower frequencies tend to be broader than those observed at higher frequencies \citep{das2020b,das2021}. The key idea is that as the frequency increases, the difference between the arrival times of the pulses of opposite circular polarisations (around a given magnetic null) will decrease, eventually becoming zero.}
    \label{fig:ideal_lightcurve}
\end{figure}

\section{Observation strategy and data analysis}\label{sec:observation}
In the ideal case of a star with an axi-symmetric dipolar magnetic field, we expect to observe a pair of oppositely circularly polarized ECME pulses around the rotational phases at which the stellar longitudinal magnetic field \bz~becomes zero (see Figure \ref{fig:ideal_lightcurve}); these rotational phases will be referred as magnetic nulls \citep[e.g.][]{leto2016}. For most of the hot magnetic stars, the magnetic field can be very well approximated as an oblique dipole \citep[e.g.][]{shultz2018,sikora2019}. If the stellar rotation axis makes an angle $i$ with the line of sight (called the `inclination angle'), and the magnetic dipole axis is inclined to the rotation axis by an angle $\delta$ (called the `obliquity), the star's \bz~curve will cross zero if $i+\delta \geq 90^\circ$. This is also the geometrical condition for the visibility of the ECME pulses. HD\,142990 has $i\approx 55^\circ$ \citep{shultz2019c} and $\delta\approx 83^\circ$ \citep{shultz2022}, and thus satisfies this criterion. Following the convention of \citet{das2019a}, we will call the null where \bz~changes from negative to positive as null 1, and the one where \bz~changes from positive to negative as null 2. We use the ephemeris reported by \citet{shultz2019_0} acording to which the null 1 corresponds to phase 0.25 and null 2 corresponds to phase 0.66.

We observed HD\,142990 around its magnetic nulls using the uGMRT band 3 ($\approx 330-470$ MHz),
and the Karl G. Jansky Very large Array (VLA) L$+$S bands (1--4 GHz, subarray mode) in the year 2019. The details of these observations are given in Table \ref{tab:targets_obs}.
The uGMRT data have single spectral window, whereas each of the VLA L and S bands are divided into sixteen spectral windows. The data were analyzed using the `Common Astronomy Software Applications' \citep[\textsc{casa},][]{mcmullin2007} following the procedure described in \citet{das2019a}. 
We supplement these data with the uGMRT band 4 ($550-900$ MHz) data for the star reported by \citet{das2019a}.

We use the IAU/IEEE convention for circular polarization throughout this paper\footnote{The uGMRT follows an opposite convention for defining RCP and LCP. We have fixed this during our post-processing of data \citep{das2020}.}.



\begin{deluxetable*}{cccccc}
\centering
\tablecaption{Observation logs for HD\,142990. HJD stands for Heliocentric Julian Day.\label{tab:targets_obs}}  
\tablehead{
Telescope & Date & HJD range & Eff. band & \multicolumn{2}{c}{Calibrator}\\
(band name) & of Obs. & $-2.45\times 10^6$ & $\Delta\nu_\mathrm{eff}$ (MHz) & Flux/bandpass & Phase 
}
\startdata
\hline
uGMRT & 2019--05--10 & $8614.32\pm 0.10$ & 334--360, 380--470 & 3C286 & J1626--298 \\
(band 3) & 2019--06--04 & $8639.32\pm 0.11$ & 344--360, 399--461 & 3C286 & J1626--298   \\
\hline
VLA & 2019--07--03 & $8667.69\pm 0.10$& 1039.5--1999.5, 2051--3947 & 3C286 & J1626--2951 \\
(L+S)& 2019--09--17 & $8744.48\pm 0.10$ & 1039.5--1999.5, 2051--3947 & 3C286 & J1626--2951 \\
\enddata
\end{deluxetable*}

\begin{deluxetable*}{ccccccc}
\centering
\tablecaption{Flux density measurements at VLA L and S bands corresponding to Figure \ref{fig:hd142990_LC_band3_band4_L_S}. `HJD1' and `HJD2' are the Heliocentric Julian Day corresponding to the beginning and end of the measurements respectively. `Pol' gives the circular polarization. The `Error' column gives the uncertainty in the flux density measurements (includes the fitting error and the image rms). $\nu_0$ represents the central frequency with a bandwidth of $\Delta\nu$. (This table is available in its entirety in machine-readable form.)\label{tab:targets_data}}  
\tablehead{
HJD1 & HJD2 & Flux density & Error & Pol & $\nu_0$ & $\Delta \nu$\\
& & (mJy) & (mJy) & & (MHz) & (MHz)
}
\startdata
\hline
2458667.676831 & 2458667.678219 & 7.28 & 0.86 & RCP & 1263.5 & 512\\
2458667.678219 & 2458667.679608 & 7.74 & 0.89 & RCP & 1263.5 & 512\\
2458667.679608 & 2458667.680997 & 8.36 & 0.93 & RCP & 1263.5 & 512\\
2458667.680997 & 2458667.682386 & 10.85 & 1.0 & RCP & 1263.5 & 512\\
2458667.682386 & 2458667.683775 & 10.83 & 1.05 & RCP & 1263.5 & 512\\
\enddata
\end{deluxetable*}

\section{Results}\label{sec:results}
We present our results obtained for uGMRT band 3, and VLA L and S bands in the following subsections.
\begin{figure}
    \centering
    \includegraphics[width=0.45\textwidth]{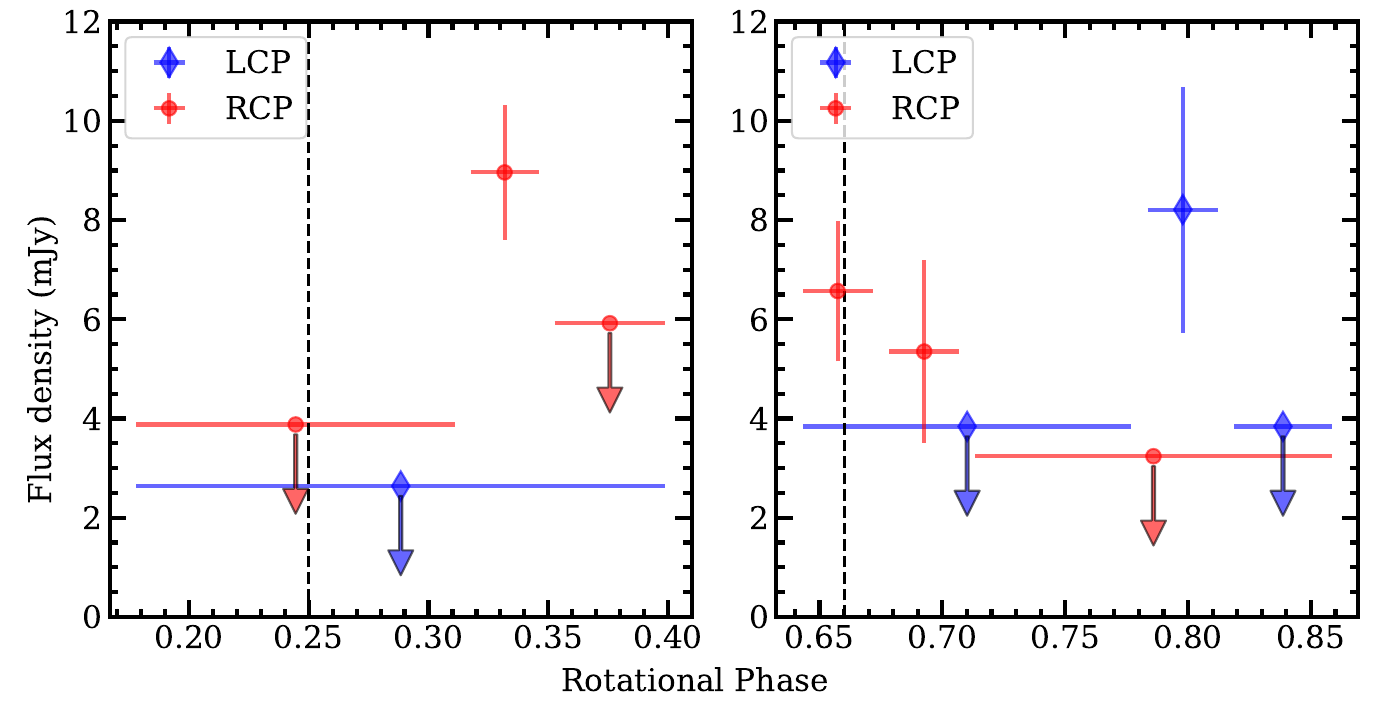}
    \caption{The lightcurves of HD\,142990 at uGMRT band 3. The vertical dashed lines mark the magnetic nulls. The arrows represent $4\sigma$ upper limits to the flux densities.}
    \label{fig:lc_band3}
\end{figure}
\subsection{Results at uGMRT band 3 ($\approx 0.4$ GHz)}\label{subsec:results_band3}
The lightcurves of HD\,142990 around its magnetic nulls at uGMRT band 3 (our lowest frequency of observation) are shown in Figure \ref{fig:lc_band3}. We did not detect the star at most rotational phases. This is due to the presence of a very bright extended source (integrated flux density $\approx 3\,\mathrm{Jy}$) less than 2 arcmin away from the target (which increases the image rms). Around null 1 (phase 0.25, left panel of Figure \ref{fig:lc_band3}), we obtained clear detection in RCP at around phase 0.33 (by averaging over 40 minutes or $\approx 0.03$ stellar rotation cycles). The corresponding flux density is $10\pm 1$ mJy. The $4\sigma$ upper limits to the flux density before and after this rotational phase are 3.9 mJy and 5.9 mJy respectively. 
In case of LCP, we did not detect the star even after averaging over the full observation session. The $4\sigma$ upper limit to the LCP flux density around null 1 comes out to be 2.6 mJy.
Around null 2 (phase 0.66, right panel of Figure \ref{fig:lc_band3}), we observed flux density enhancement in both circular polarizations. While it appears that we covered the RCP pulse only partially, the LCP pulse is covered entirely by our observing phase-window. The peak flux densities at RCP and LCP are $7\pm1$ mJy and $8\pm2$ mJy respectively.

We conclude that there are clear indications of RCP enhancement following the rotational phase of null 1, and enhancements in both circular polarizations around the other null (null 2). The observed sequence of enhancements at this frequency is consistent with X-mode ECME. 
Note that the star has been observed to produce highly circularly polarized emission at 200 MHz around null 2 \citep[][Emil Lenc, private commnucication]{lenc2018}. Thus the ECME spectra for pulses at null 2 clearly extend below 400 MHz, but the same is not clear for the pulses around null 1.

\begin{figure}
    \centering
    \includegraphics[width=0.48\textwidth]{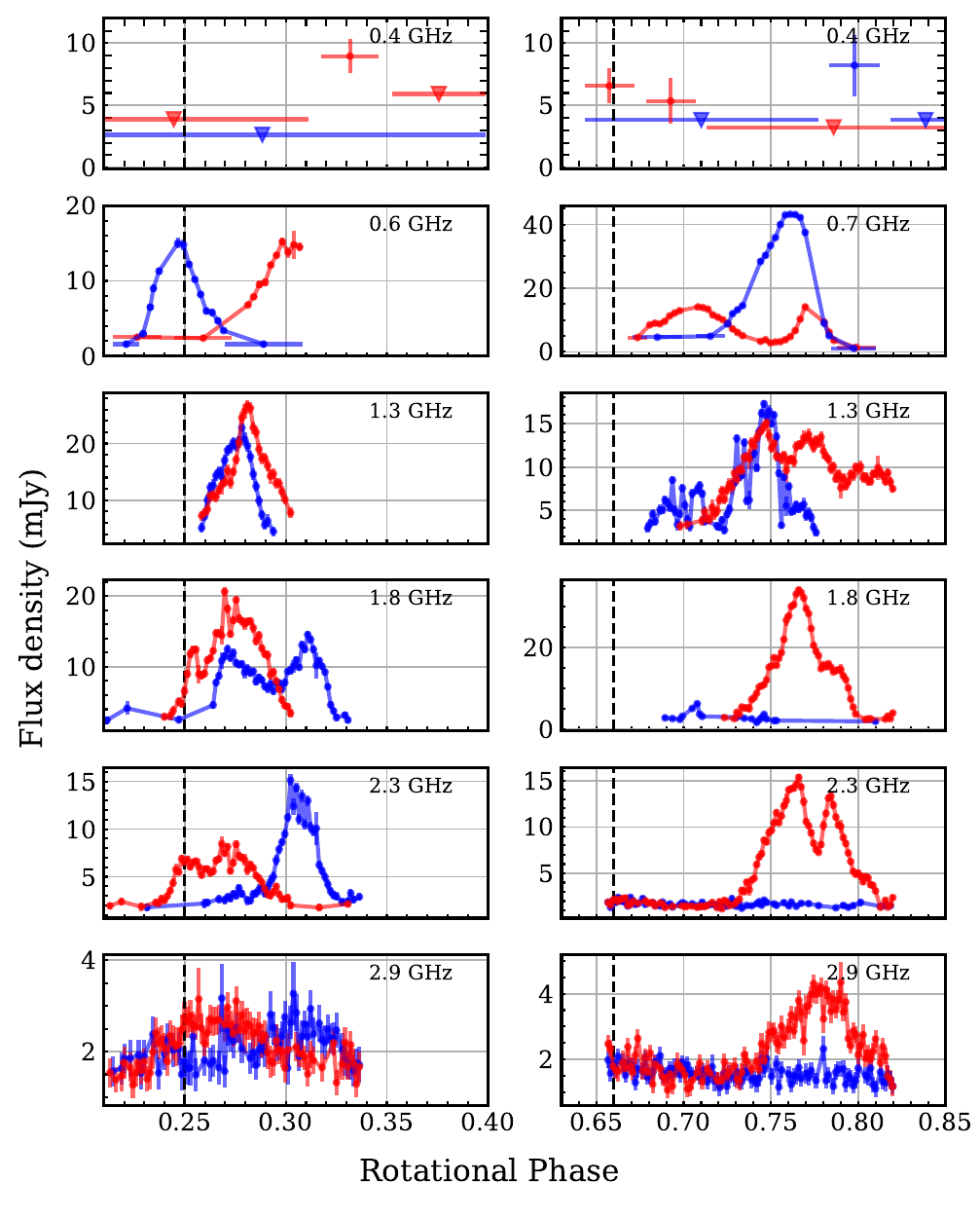}
    \caption{The lightcurves of HD\,142990 over the frequency range of 0.4--4 GHz. The triangles in the top panels represent $4\sigma$ upper limits to the flux density. The red and blue markers correspond to RCP and LCP, respectively. The vertical dashed lines mark the rotational phases of the magnetic nulls. Note that the data used in the second panels are obtained from \citet{das2019a}.}
    \label{fig:hd142990_LC_band3_band4_L_S}
\end{figure}

\subsection{Results at VLA L and S bands (1--4 GHz)}\label{subsec:results_Lband}
We divided the VLA L band (1--2 GHz) into two sub-bands, the one spanning the frequency range $\approx 1.0-1.5$ GHz, and the other spanning the range $\approx 1.5-2.0$ GHz. As will be described in a subsequent section, the ECME spectra for all four pulses exhibit cut-off at around 3 GHz (\S\ref{subsec:ecme_spectra}). We hence consider only the first half of the S band (2--3 GHz) to investigate the frequency dependence of ECME pulse-profiles. Table \ref{tab:targets_data} gives the corresponding flux density measurements. In Figure \ref{fig:hd142990_LC_band3_band4_L_S} (third panels onwards), we show the lightcurves around the two magnetic nulls at four frequency bins that approximately span the frequency ranges of 1--1.5 GHz, 1.5--2.0 GHz, 2.0--2.5 GHz and 2.5--3.0 GHz, along with the lightcurves at 0.4 and 0.7 GHz. We find the pulses to be extremely peculiar in many aspects which are described below.

\subsubsection{ECME pulses around Null 1}\label{subsubsec:null1}
We first consider the pulses observed around null 1 (left panels of Figure \ref{fig:hd142990_LC_band3_band4_L_S}). According to the ideal scenario of frequency dependence of ECME pulses \citep[][see Figure \ref{fig:ideal_lightcurve}]{trigilio2011,leto2016,das2020a}, we expect that with the increase of the observing frequency, the difference between the arrival times of RCP and LCP pulses (around a given magnetic null) will decrease, eventually they will overlap with each other (bottom panels of Figure \ref{fig:ideal_lightcurve}). As a result, above that frequency, we will observe a pulse with nearly zero circular polarization. We also expect that the shift of the RCP and LCP pulses in the phase-axis will be in opposite directions\footnote{If $\nu_1$ and $\nu_2$ are two frequencies such that $\nu_1<\nu_2$, if the LCP pulse at frequency $\nu_2$ arrives \textbf{after} the LCP pulse at frequency $\nu_1$, the RCP pulse at $\nu_2$ will arrive \textbf{ahead} of the RCP pulse at $\nu_1$ \citep[e.g.][]{leto2016}. Also see Figure 8 of \citet{das2020a}.} (as illustrated in Figure \ref{fig:ideal_lightcurve}). In our case, however, we find that while up to a frequency of $\approx 1.3$ GHz, the pulses behave as expected, beyond that frequency, the behavior (especially for the LCP pulse) is totally inconsistent with the ideal scenario. The RCP pulse continues its shift towards `left' (phase-axis) over $\approx0.4-2$ GHz, and beyond that, it roughly appears over the same range of rotational phases (see Figure \ref{fig:hd142990_LC_band3_band4_L_S}). In case of the LCP pulse, from 0.6 to 1.3 GHz, it shifts towards right (Figure \ref{fig:hd142990_LC_band3_band4_L_S}, thus opposite to the direction of the shift of the RCP pulse). At 1.8 GHz, it assumes a double-peaked morphology, with the first peak appearing at approximately over the same range of rotational phases as that at 1.3 GHz, but the second peak arrives much later in time (after the RCP pulse). At 2.3 GHz, we get back two clearly separated RCP and LCP pulse, but now their sequence of arrival is opposite to what was observed at lower frequency (e.g. at 0.6 GHz, second left panel of Figure \ref{fig:hd142990_LC_band3_band4_L_S}). The pulses nearly disappear at around 3 GHz. It is also interesting to note that the second peak of the LCP pulse appears to shift towards left (phase-axis) between 1.8 and 2.3 GHz, which is opposite to the direction along which the lone LCP pulse shifts over 0.6--1.3 GHz.

\subsubsection{ECME pulses around Null 2}\label{subsubsec:null2}
\citet{das2019a} noted that the RCP \footnote{\citet{das2019a} reported that the LCP pulse near null 2 exhibits double-peaked morphology. However, the convention used by them was opposite to the IAU/IEEE convention \citep{das2020_erratum1}. Thus, under the IAU/IEEE convention, the polarization of that pulse is RCP.} pulse around null 2 exhibits a peculiar double-peaked profile, which, at that time, was not reported for any other hot magnetic star. Here we find that till a frequency of around 1.8 GHz, the sub-pulses come closer to each other, with an overall shift towards right (in time/phase axis), and they form one single RCP pulse at $\approx 1.8$ GHz. However, at 2.3 GHz, once again, we get the double-peaked structure, which again becomes a single pulse at $\approx 3$ GHz. The LCP pulse, on the other hand, exhibits relatively simple changes with frequency. Its peak flux density quickly decreases within the L band, and becomes almost undetectable in the S band. Between 0.7 and 1.3 GHz, it shifts towards left (thus opposite to the direct of shift for the corresponding RCP pulse).

For both magnetic nulls, the pulses do not lie symmetrically about the magnetic null phases, which has also been observed for many other MRPs \citep[e.g. CU\,Vir, HD\,133880,][etc.]{das2021,das2020b}. This could be due to limitation of the adopted ephemeris, or the fact that the stellar magnetic fields are not perfectly axi-symmetric dipolar.

We provide lightcurves per spectral window of L and S bands in the appendix (Figures \ref{fig:hd142990_L_full_lc} and \ref{fig:hd142990_S_full_lc}) which clearly show the evolution of pulses with frequencies. These lightcurves are used to obtain the peak flux density spectra of the ECME pulses described in the next subsection.

\subsection{The peak flux density spectra and ECME upper cut-offs}\label{subsec:ecme_spectra}
\begin{figure*}
\centering
\includegraphics[width=0.98\textwidth]{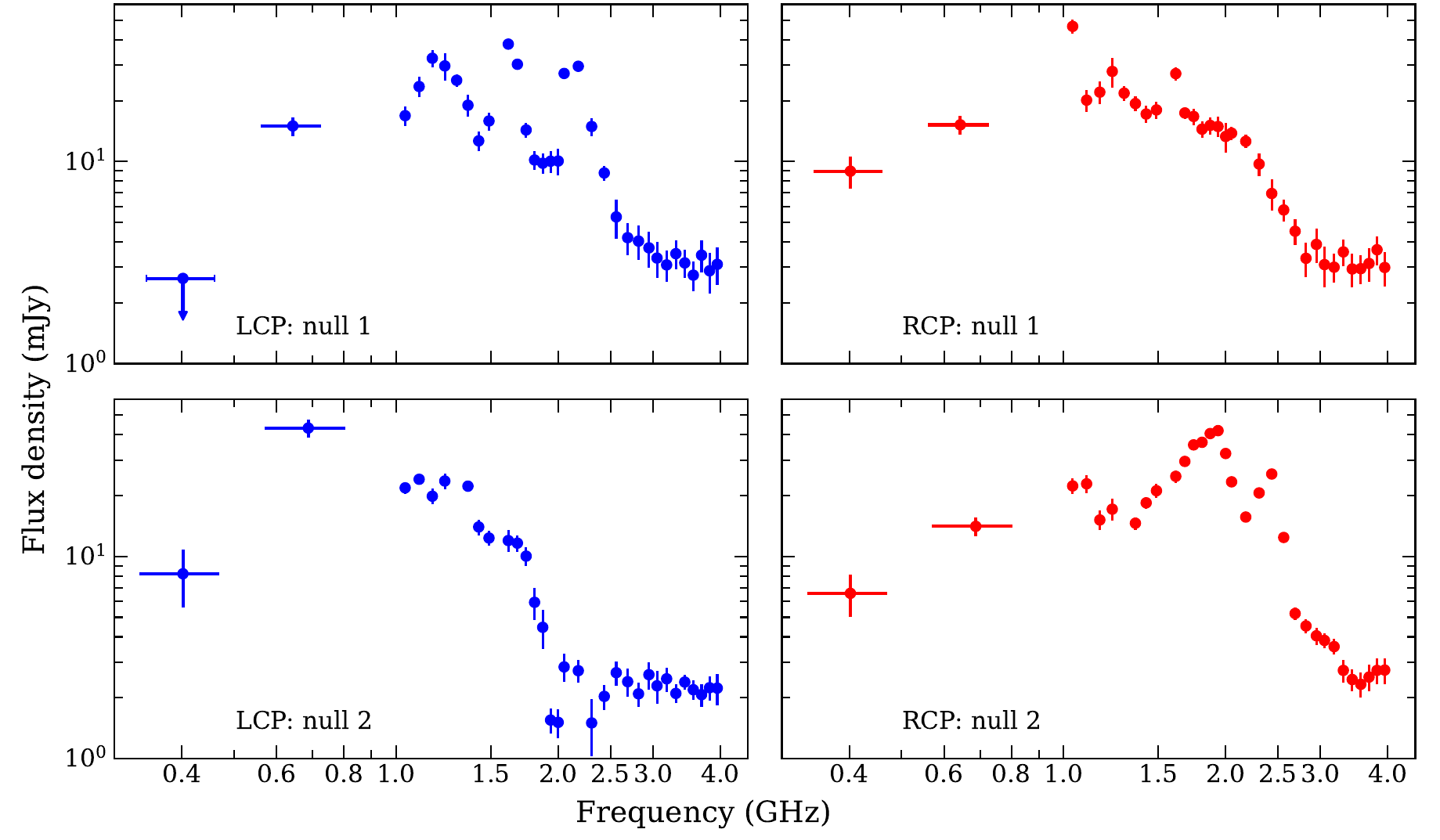}
\caption{The peak flux density spectra for the ECME pulses from HD\,142990 near the two magnetic nulls over our entire frequency range of observations.}
\label{fig:hd142990_spectra}
\end{figure*}

We show the peak flux density spectra in Figure \ref{fig:hd142990_spectra}. Like the other three MRPs that have been studied over the same range of frequencies, the spectra for the RCP and LCP pulses around the two nulls are non-identical. The spectral shapes are extremely peculiar and unique in the sense that there are local maxima and minima between 1--4 GHz (e.g., see the upper left panel of Figure \ref{fig:hd142990_spectra}). These features are very prominent for the LCP pulse near null 1, and the RCP pulse near null 2, and the least prominent (probably absent) for the LCP pulse near null 2. The possible reason(s) for this feature will be discussed in the next section.

\begin{figure*}
\centering
\includegraphics[width=0.98\textwidth]{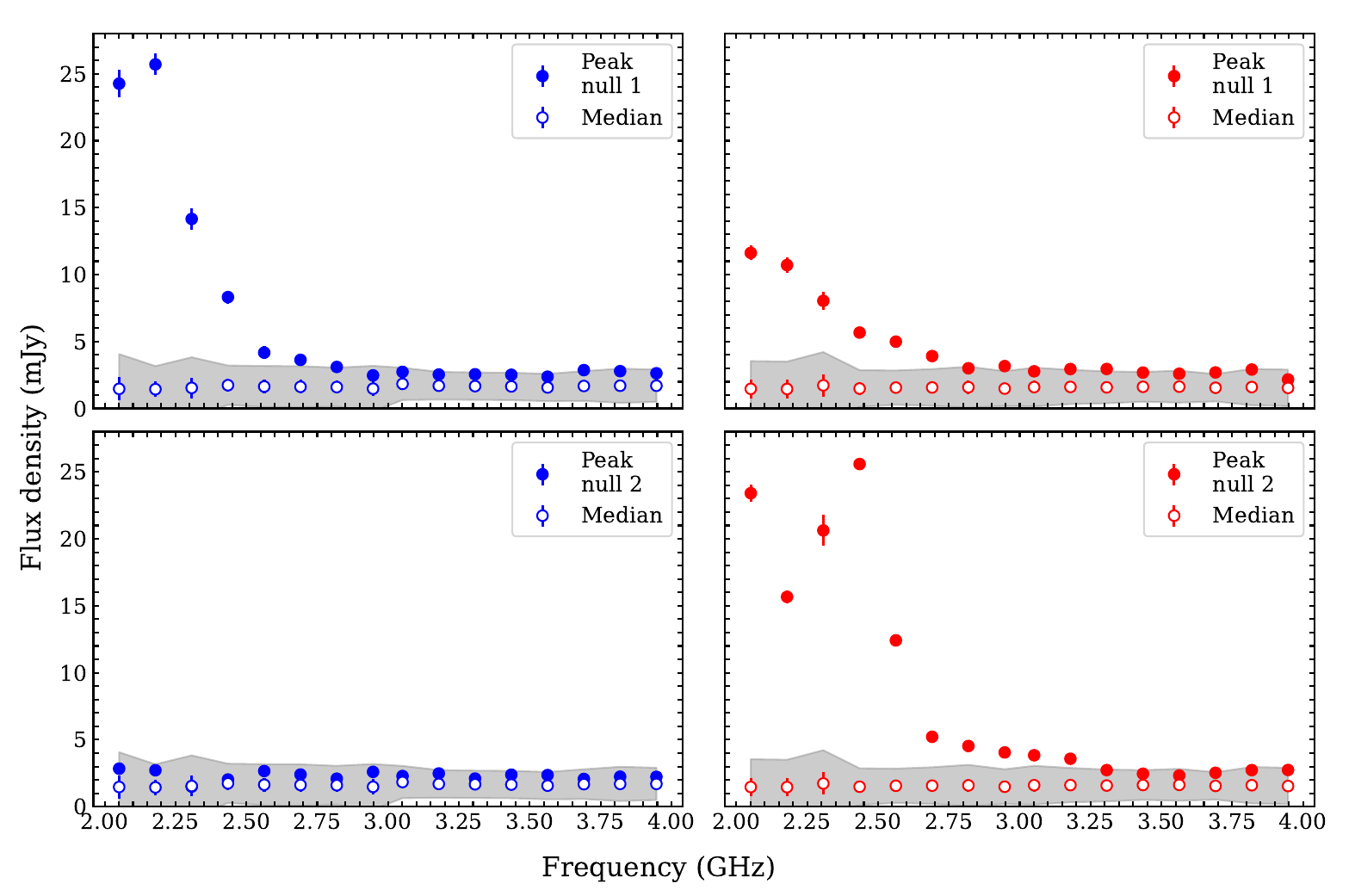}
\caption{The peak flux density spectra (filled markers) for the ECME pulses from HD\,142990 near the two magnetic nulls over 2--4 GHz (VLA S band). Null 1 is phase 0.25 where \bz \,changes its sign from negative to positive and null 2 is phase 0.66 where \bz\,changes its sign from positive to negative. The averaging time is 6 minutes. Also shown are the basal flux density spectra (unfilled markers). The grey shaded area surrounding them represent $3\sigma$ deviation from the basal spectra.}
\label{fig:hd142990_cutoff}
\end{figure*}

In order to locate the upper cut-off frequencies for the four pulses, we employ a similar strategy as was used in the case of HD\,133880 \citep{das2020b}. In case of null 1, the rotational phase range, over which we search for maximum flux density is 0.21--0.34, and that for null 2 is 0.65--0.82. For the basal flux densities, we averaged over the rotational phase ranges 0.18--0.21 and 0.34--0.37 (where there are no enhancements). This was done as over these rotational phase ranges, we could not detect the target with a time resolution comparable to that employed for the rotational phase ranges containing the enhancements. With just one data point for basal flux density (per spectral window), it is not possible to follow the procedure used in the case of HD\,133880 for this aspect. Hence to obtain a meaningful comparison between the peak flux density data points (with an averaging time of 6 minutes, Figure \ref{fig:hd142990_cutoff}) and the basal flux density data points (with an averaging time of 97 minutes), we scaled up the errorbars of the basal flux density measurements by a factor $\sqrt{97/6}\approx 4$. We define the upper cut-off frequency as the minimum frequency at which the peak flux density becomes consistent with the basal flux density within $3\sigma$ uncertainty of the latter \citep{das2020b}.
We find that for the LCP pulses, the cut-off frequencies are $ 2.8\,\mathrm{GHz}$ near null 1 and $\approx  2\,\mathrm{GHz}$ near null 2 (Figure \ref{fig:hd142990_spectra}).
For RCP pulses, the cut-off frequency is at $ 2.8$ GHz near null 1 and at $3.3$ GHz near null 2.

\section{Discussion}\label{sec:discussion}
\begin{figure}
    \centering
    \includegraphics[width=0.45\textwidth]{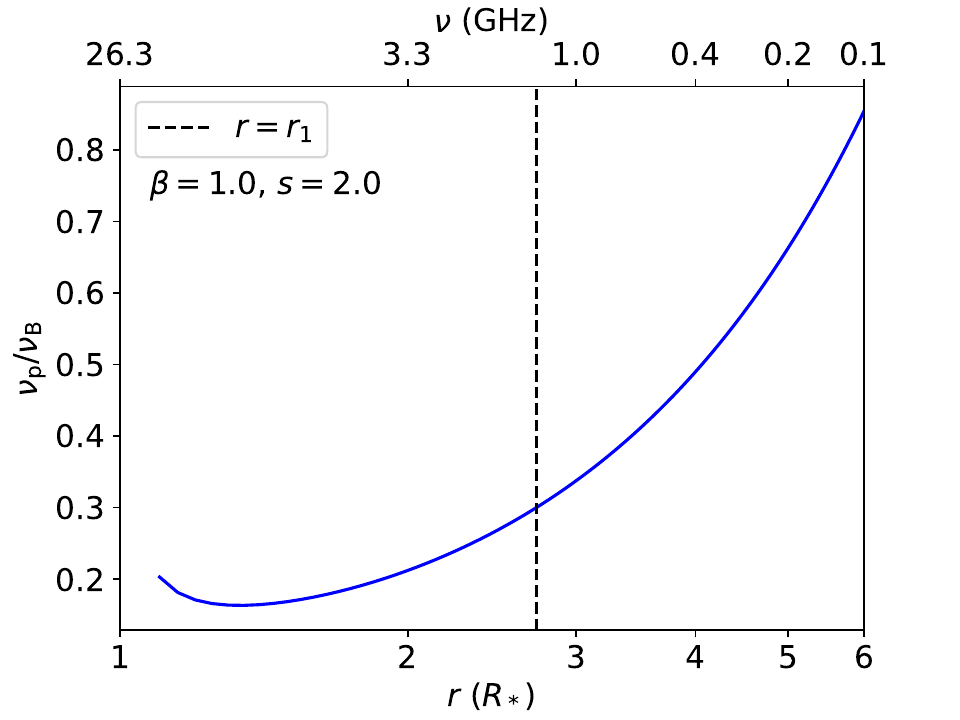}
    \caption{The radial variation of the ratio of plasma frequency to electron gyrofrequency $\nu_\mathrm{p}/\nu_\mathrm{B}$ under the assumption that at 1.3 GHz ($r=r_1$), this ratio equals 0.3. $s$ is the harmonic number and $\beta$ is the index of the velocity law. See \S\ref{sec:discussion} for details.}
    \label{fig:nup_nuB}
\end{figure}

Out of the 18 MRPs discovered so far, only four (including HD\,142990) have been observed over sub-GHz to GHz frequencies. Although none of these stars exhibit ideal behavior as depicted in Figure \ref{fig:ideal_lightcurve}, HD\,133880 and HD\,35298, nevertheless, conform to the key characteristic that as the frequency increases, the separation between oppositely circularly polarized pulses near a given magnetic null decreases \citep[see Figure 2 and 1 of][respectively]{das2020b, das2022b}. CU\,Vir \citep[the first MRP discovered,][]{trigilio2000}, on the other hand, was found to be completely inconsistent with the expectation from the ideal scenario in terms of its radio pulses, especially at sub-GHz frequencies \citep[see Figure 12 of][]{das2021}. Note that this is the only MRP which was observed for its one full rotation cycle over 0.4--4 GHz, and it led to the surprising discovery that persistent radio pulses can appear at rotational phases significantly offset from the magnetic nulls \citep{das2021}.
The properties of the radio pulses from HD\,142990 is more complex than those observed from HD\,35298 or HD\,133880. They also differ from those observed from CU\,Vir in the sense that for HD\,142990, the pulse-properties become complex at GHz frequencies, whereas for the former, it was the sub-GHz frequency range where non-trivial features were observed. This could be due to the differences in their polar magnetic field strengths \citep[HD\,142990 has a stronger magnetic field than that of CU\,Vir,][]{shultz2019c,trigilio2000} as the ECME frequencies are directly proportional to the magnetic field strengths at the emission sites. However, it is also difficult to compare the two as unlike CU\,Vir, HD\,142990 has not been observed for its one full rotation cycle. Thus, future studies should aim at acquiring lightcurves over full rotation cycles so as to examine whether this object also exhibits radio pulses away from magnetic nulls.

A unique characteristic observed from HD\,142990 is the apparent reversal of the sequence of arrival of the oppositely circularly polarized pulses around null 1 (left panels in Figures \ref{fig:ideal_lightcurve} and \ref{fig:hd142990_LC_band3_band4_L_S}). The importance of this observation lies in the fact that the sequence is often used to identify the magneto-ionic mode of emission \citep[see Figure 1 of][]{das2019a}, which, in turn, is used to estimate the plasma density at the emission site \citep[e.g.][]{leto2019,das2019b}. The latter relies on the theoretical idea that the extra-ordinary (X-) mode dominates when the ratio between the plasma frequency to the electron gyrofrequency $\nu_\mathrm{p}/\nu_\mathrm{B}\lesssim 0.3$, and the ordinary (O-) mode dominates for $\nu_\mathrm{p}/\nu_\mathrm{B}\gtrsim 0.3$ \citep{sharma1984,melrose1984}. 
Note that the critical value (defining the `mode-transition') of $\nu_\mathrm{p}/\nu_\mathrm{B}$ is a function of electron energies, and could be higher for hot magnetic stars if the electrons are more energetic than solar and planetary cases \citep[for which the above critical values were estimated][]{melrose1984, sharma1984}. But the general idea is that below a critical value of this ratio, the X-mode dominates, and above that, the O-mode dominates as the cut-off for the former becomes higher than the emission frequency.
In case of X-mode emission, we expect to see an RCP followed by LCP pulse near null 1, and an LCP followed by RCP around null 2, whereas for the O-mode, it is the opposite \citep[e.g. see Figure 1 of][]{das2019a}.

\begin{figure*}
    \centering
    \includegraphics[width=0.45\textwidth]{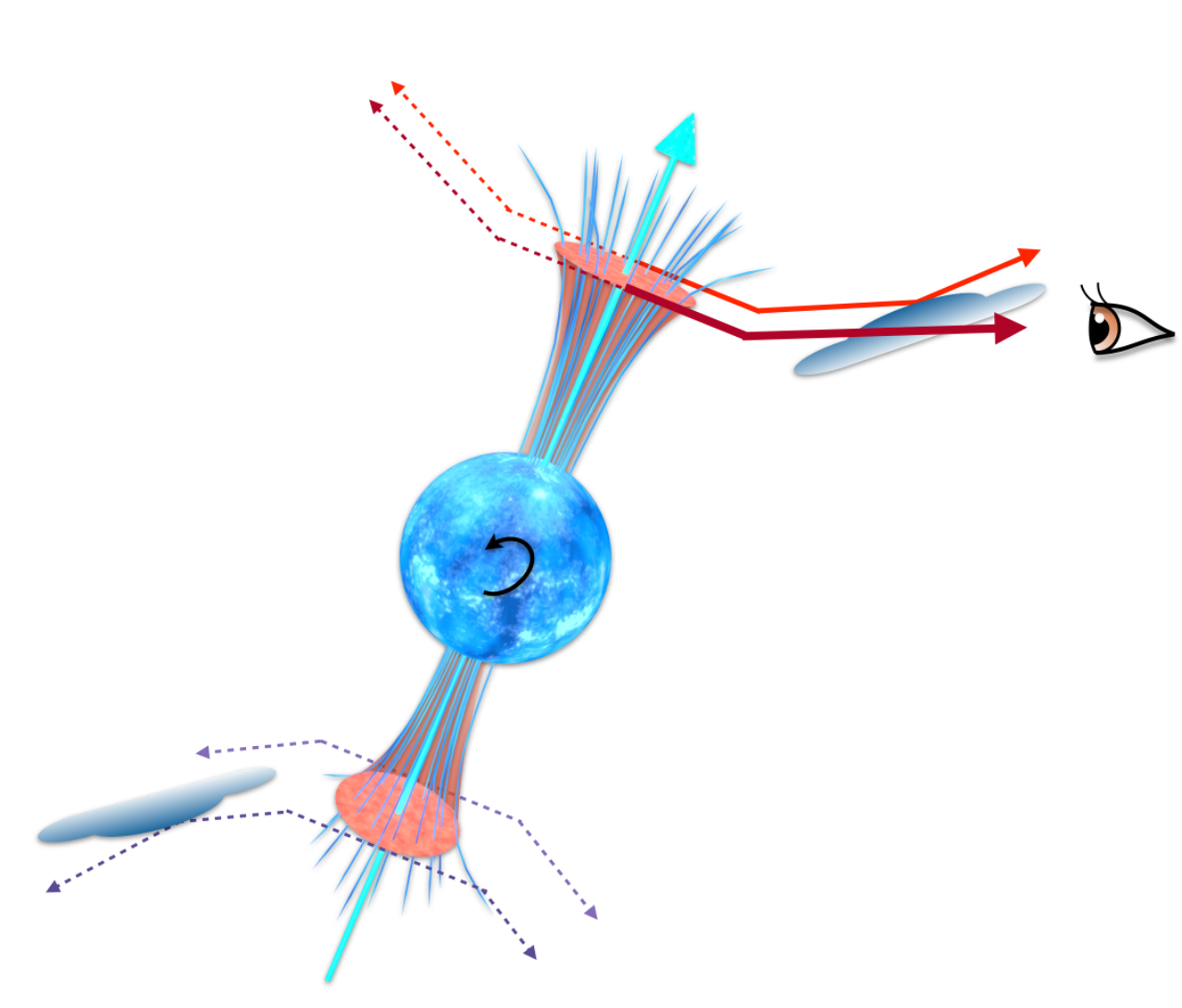}
    \includegraphics[width=0.45\textwidth]{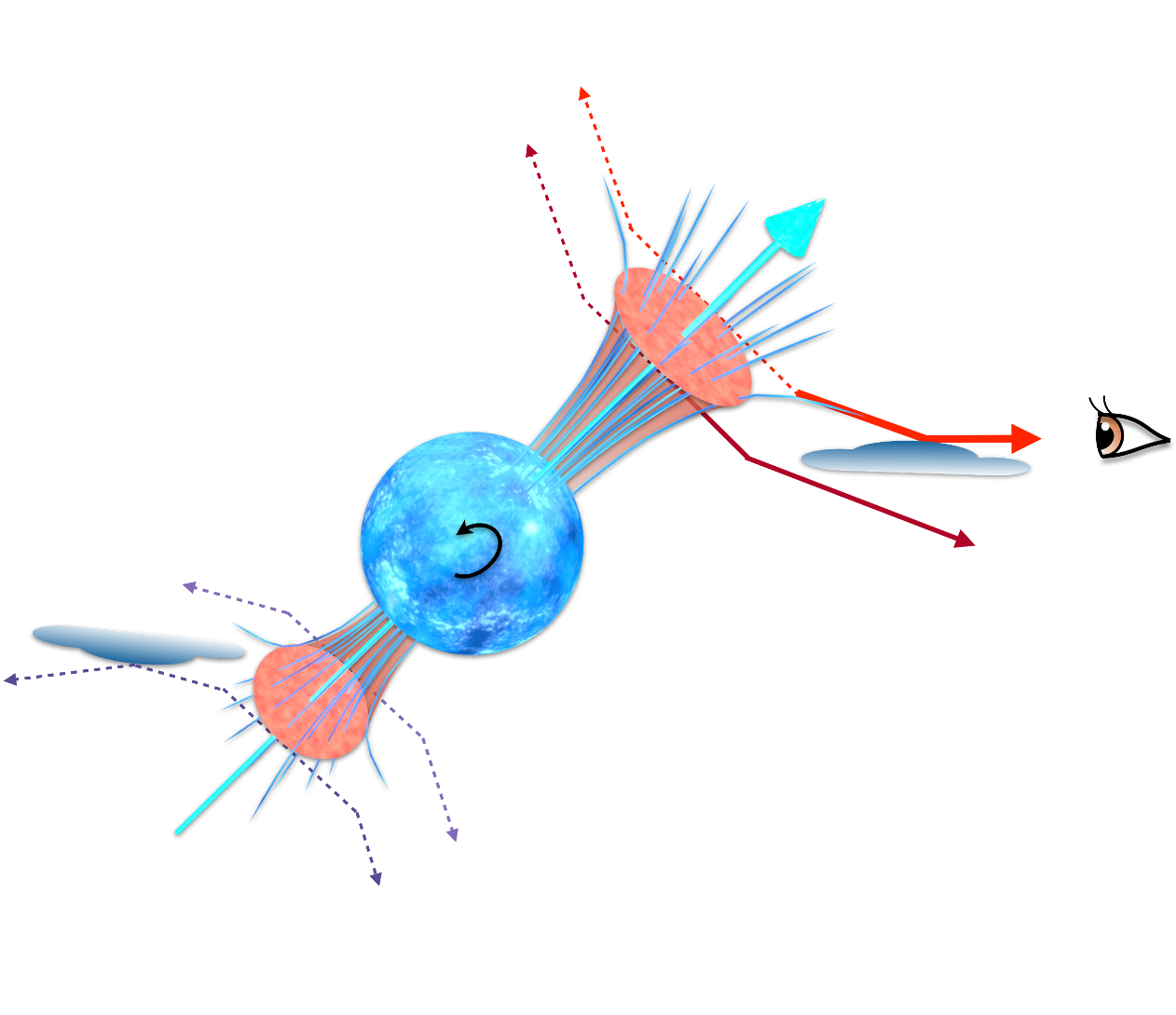}
    \caption{A cartoon diagram illustrating the propagation effects suffered by ECME in a magnetosphere which has a plasma distribution not symmetric to the magnetic axis (shown by the cyan arrow). For hot magnetic stars with an axi-symmetric magnetic field, ECME is produced in auroral rings, and emitted tangential to it \citep{mutel2008,trigilio2011}.}
    \label{fig:auroral_emission_sites}
\end{figure*}

Around null 1, the observed pulse arrival sequence at 0.6 GHz is consistent with X-mode emission (for which we expect to see the LCP followed by the RCP pulse around null 1). If we now simply consider the pulses at 2.3 GHz (Figure \ref{fig:hd142990_LC_band3_band4_L_S}, fifth left panel), without considering the evolution in the intermediate frequencies, the straight-forward inference will be that the magneto-ionic mode is ordinary at 2.3 GHz. This gives the impression that there is a magneto-ionic mode transition between 0.6 and 2.3 GHz. To have a transition from X-mode to O-mode with increasing frequency, which is equivalent to decreasing radial distance $r$ (higher frequency originates closer to the star where the magnetic field is stronger, and vice-versa), we will require $\nu_\mathrm{p}/\nu_\mathrm{B}$ to increase as we go closer to the star.
This is, however, opposite to what we expect in stellar magnetospheres as $\nu_\mathrm{B}$ increases much faster with decreasing radial distance than that for $\nu_\mathrm{p}$ \citep{trigilio2008}. We can also check it quantitatively using Eq. 4 of \citet{das2022b}, with the constraint that $\nu_\mathrm{p}/\nu_\mathrm{B}\approx 0.3$ at around 1.3 GHz (where RCP and LCP pulses overlap around null 1). We show the radial variation of $\nu_\mathrm{p}/\nu_\mathrm{B}$ in Figure \ref{fig:nup_nuB} considering emission at the second harmonic, and $\beta=1$ \citep[$\beta$ is the index of the stellar wind velocity law,][]{castor1975}. Here $r_1$ is the radial distance at which ECME at 1.3 GHz is produced (hence $\nu_\mathrm{p}/\nu_\mathrm{B}=0.3$ at $r=r_1$). For our hypothesis of mode-transition from X to O-mode with decreasing $r$, we need $\nu_\mathrm{p}/\nu_\mathrm{B}>0.3$ for $r<r_1$, which is clearly not the case. The same result is obtained for emission at the fundamental harmonic. 
Note that the shape of the curve is independent of the assumed critical value of $\nu_\mathrm{p}/\nu_\mathrm{B}$. Thus, Figure \ref{fig:nup_nuB} demonstrates that it is unlikely to have ordinary mode ECME at higher frequencies.

The other scenario for giving rise to peculiar frequency dependence is propagation effect in the stellar magnetosphere. 
\citet{das2020a} provided a numerical framework to take into account of refraction experienced by the emission while passing through the inner magnetospheric plasma for any kind of plasma distribution. Prior to that, to predict the effects of the magnetospheric plasma on the ECME pulses, the stellar magnetosphere (bounded by the largest closed magnetic field line) was effectively approximated as a region with constant plasma density \citep{trigilio2011}. This gives rise to the frequency dependence similar to those shown in Figure \ref{fig:ideal_lightcurve}.
\citet{das2020a} considered more realistic magnetospheric plasma distributions inspired from those predicted by the highly successful `Rigidly Rotating Magnetosphere' \citep[RRM,][]{townsend2005} model. According to the RRM model, the magnetospheric plasma distributes itself around the magnetic equator in an azimuthally symmetric manner when the magnetic axis is nearly aligned with the rotation axis. On the other hand, as the misalignment between the two axes increases, the distribution becomes significantly asymmetric about the magnetic axis, and also it no longer remains confined to the magnetic equator.
The key result obtained by \citet{das2020a} was that when most of the plasma are distributed around the magnetic equatorial plane in an azimuthally symmetric fashion (the case of small obliquity), the net result is qualitatively the same as that predicted assuming that the inner magnetosphere has constant plasma density. However, the pulse-properties become significantly complicated when plasma accumulates away from the magnetic equator, and without maintaining the azimuthal symmetry about the magnetic axis (the case of large obliquity). 
Two non-trivial characteristic that emerged in this case
are appearance of secondary pulses at certain frequencies, and reversal of the sequence of arrival of pulses at higher frequencies \citep{das2020a}. Note that these non-ideal features appeared despite considering an axi-symmetric dipolar magnetic field.

In case of HD\,142990, the magnetic axis is nearly at right angle to the rotation axis \citep{shultz2022}, suggesting that propagation effects could play an important role in defining the observed pulse characteristics.
There is, however, another requirement for this scenario to be relevant, which is that the magnetosphere must be dense enough so as to be able to significantly bend the radiations at GHz frequencies. According to the simulations performed by \citet{das2020a}, a plasma density $\sim 10^{11}\,\mathrm{cm^{-3}}$ is sufficient to cause the reversal in the arrival sequence of the oppositely circularly polarized pulses (see their Figures 11 and 12). In order to examine whether the magnetosphere of HD\,142990 can have plasma with density $\sim10^{11}\,\mathrm{cm^{-3}}$, we estimate the maximum possible magnetospheric plasma density using the following equations \citep{townsend2005,owocki2020}:
\begin{align}
    \rho_\mathrm{max}&=\frac{\sigma_\mathrm{K}}{\sqrt{\pi}h_\mathrm{K}}\label{eq:density_eq}\\
    \sigma_\mathrm{K}&=\frac{0.3B_\mathrm{K}^2}{4\pi g_\mathrm{K}},\,g_\mathrm{K}=\frac{GM_*}{R_\mathrm{K}^2}\nonumber \\
    h_\mathrm{K}&=\sqrt{2r_\mathrm{K}}\epsilon_*R_\mathrm{K}\nonumber
\end{align}
where $B_\mathrm{K}$ is the magnetic field strength at the Kepler radius $R_\mathrm{K}$ at the magnetic equator, $M_*$ is the stellar mass, $r_\mathrm{K}=R_\mathrm{K}/R_*$, where $R_*$ is the stellar radius, $\epsilon_*$ is the ratio of thermal to gravitational binding energy at the stellar surface and is $\sim 0.001$ for early-type stars \citep{townsend2005}. Using the stellar parameters for HD\,142990 \citep{shultz2019c}, we obtain $\rho_\mathrm{max}\approx 7.3\times 10^{-11}\,\mathrm{g/cm^{3}}$. The corresponding number density is obtained by dividing $\rho_\mathrm{max}$ by proton mass, which comes out to be $\sim 10^{13}\,\mathrm{cm^{-3}}$.
Thus, propagation effect is a promising scenario to explain the observed phenomenon.

We now consider the other peculiarities, apart from the reversal of the arrival sequence, exhibited by the radio pulses from HD\,142990.
We find that near null 1, the arrival phases of the LCP pulse exhibit large amount of shift as a function of frequency (peak shifts from phase 0.27 to phase 0.31 from 1 to 2 GHz, Figure \ref{fig:hd142990_L_full_lc}), as compared to that for the peak of the RCP pulse, which merely shifts from 0.28 phase at 1.0 GHz to 0.27 phase at 2.0 GHz. By looking at the left panels of Figure \ref{fig:hd142990_L_full_lc}, we find that this is due to the fact that a secondary LCP pulse starts to appear at around 1.7 GHz over the phase range 0.30--0.33. The original LCP pulse over 0.25--0.30 phases becomes much weaker beyond this frequency, and at 2 GHz, the `secondary' LCP pulse becomes the only LCP pulse observable around null 1. 
As mentioned already, the appearance of secondary pulses over a certain frequency range can also be explained considering propagation effects. Following \citet{das2020a}, we propose that the different pulses likely originate at different emission sites with different magnetic longitudes.
If the magnetosphere has a plasma distribution symmetric about the magnetic axis, the radiation produced at a given magnetic latitude (but with different magnetic longitudes), will experience identical plasma conditions as they pass through the magnetosphere. Thus, rays that are parallel at the time of emission, will remain parallel when they emerge out of the stellar magnetosphere. However, for the highly misaligned rotation and magnetic axes of HD\,142990, its magnetosphere is unlikely to have such symmetric plasma conditions. As a result, the rays produced at emission sites located along different magnetic longitudes, will
experience different plasma densities along their way to the observer, and will suffer different amounts of deviations, leading to their different arrival times, manifested as separated pulses (see Figure \ref{fig:auroral_emission_sites}).
Note that our observation of double peaked pulse before observing the reversal of pulse-arrival sequence is consistent with the prediction of \citet{das2020a}.

Although not as evident as that for null 1, there is a hint of reversal of arrival sequence for pulses near null 2 as well. At our lowest frequency of observation, our observation suggests that the RCP pulse arrives ahead of the LCP pulse (see Figure \ref{fig:lc_band3}). If we now consider the pulses at 1.6 GHz (right panels of Figure \ref{fig:hd142990_L_full_lc}), our data suggest that the (much fainter) LCP pulse arrives ahead of the RCP pulse. This is difficult to appreciate since the LCP pulse near null 2 is extremely weak above $\approx$ 1.4 GHz. It is interesting to note that the frequency dependence exhibited by the RCP pulse near null 2 is qualitatively very similar to that for the LCP pulse near null 1 in the sense that it also potentially has a single-peaked profile at lower frequencies (0.4 GHz), beyond a certain frequency ($\approx0.6$ GHz), it assumes a multi-peaked profile, eventually causing an apparent reversal of pulse-arrival sequence at a higher frequency ($\approx 1.4$ GHz).

The qualitative similarity between the frequency dependence of the LCP pulse near null 1, and the RCP pulse near null 2 can be seen more clearly in Figure \ref{fig:hd142990_spectra} which shows the peak flux density spectra. Both pulses have spectra with local maxima between 1--4 GHz. 
Note that although ECME pulses from MRPs have been reported to exhibit time-variability \citep[e.g.][]{trigilio2011,das2021}, the individual spectra over 1--4 GHz are robust against this possibility since they were acquired simultaneously using sub-array mode (see \S\ref{sec:observation}). While it will still be interesting to examine how (or whether) the spectra evolve with time, the only time such a study was reported was for CU\,Vir by \citet{das2021}, and they did not observe any significant change in spectral shapes for their two epochs of observations separated by nearly a year.

Under the scenario that the spectra shown in Figure \ref{fig:hd142990_spectra} represent stable characteristics of the system, we propose that these features indicate that multiple pulses, that originate at different emission sites (characterized by different magnetic longitudes) in the stellar magnetosphere, are present over this frequency range. While from the lightcurves alone, we can clearly see two different LCP pulses near null 1, the spectrum suggest that there are actually three, with two of them appearing over similar rotational phase ranges. For the RCP pulse near null 2, both lightcurves (Figures \ref{fig:hd142990_L_full_lc} and \ref{fig:hd142990_S_full_lc}) and spectrum suggest that at least three different pulses are observable over 1--4 GHz. Similar features can also be seen in the spectra for the RCP pulse near null 1 and LCP pulse near null 2, but at a much weaker level.

An alternative to the propagation effect scenario is complex magnetic field topology (instead of large-scale, nearly dipolar topology of the stellar magnetic field). However, this scenario requires a significant deviation from the currently accepted ideas about hot stars' magnetospheres. In contrast, propagation effects provide an explanation to the observed peculiarities that emerges naturally from the current ideas about how magnetospheric plasma is accumulated surrounding these stars.

With an upper cut-off at $\approx 3$ GHz, and a polar magnetic field strength of $\approx 4.7$ kG \citep[][corresponding electron gyrofrequency is 13 GHz]{shultz2019c}, HD\,142990 reinforces the idea that premature upper cut-off frequency is a general characteristic of ECME produced by hot magnetic stars. In the future, it will be extremely important to investigate if the same is true for other objects with large-scale, ordered magnetic fields (e.g. UCDs), since, ECME upper cut-offs are often used as a measure of maximum surface magnetic field strengths in these cooler objects \citep[e.g.][]{kao2023}.

\section{Conclusion}\label{sec:conclusion}
In this paper, we present observation of HD\,142990 around its magnetic nulls, over 0.4--4.0 GHz using the uGMRT and the VLA, with an aim to investigate the frequency dependence of ECME pulse profiles. Other than this star, only three other MRPs have been studied in the same fashion. 
HD\,142990 exhibits certain common properties, such as different spectra for the pulses observed at different rotational phases (orientation dependence of ECME pulses, Figure \ref{fig:hd142990_spectra}), premature upper cut-off ($\lesssim 3.3$ GHz) and asymmetric arrival phases about the magnetic nulls (e.g. Figure \ref{fig:hd142990_LC_band3_band4_L_S}). In addition to these, it exhibits the unique property of reversal of the sequence of arrival of oppositely circularly polarized pulses within 1--4 GHz, a phenomenon that has already been predicted based on numerical simulation of ECME lightcurves for stars that have magnetospheres with extremely complex plasma distributions \citep{das2020a}, but has never been observed before. With an obliquity of $\approx 90^\circ$, HD\,142990 is extremely likely to possess such complex plasma distribution in its magnetosphere \citep{townsend2005}, and thus our result, once again, demonstrates that the inner magnetospheric plasma plays a key role in defining the observed characteristics of ECME. One of the most important takeaways of this work is that inferring the magneto-ionic mode based on ECME pulse arrival sequence over a narrow frequency range is not trivial. Our work also raises the need to observe more MRPs with varying physical properties over wide frequency ranges, so that we can eventually understand the seemingly non-identical behavior of these stars. For example, to confirm whether or not the obliquity parameter is important to create complex frequency dependence, it will be interesting to observe MRPs with low obliquities. In addition, it is now evident that ECME pulses can appear significantly offset from the magnetic nulls \citep[e.g. CU\,Vir,][]{das2021}, hence, it will be more fruitful to acquire observation of MRPs over full rotation cycles instead of observing them only around their magnetic nulls. Along with observation, it will also be important to attempt to perform numerical simulation of ECME lightcurves \citep[e.g. using the framework of][]{das2020a} considering magnetospheric plasma density distribution predicted by existing theoretical framework \citep[e.g.][]{townsend2005, berry2022}, to examine the extent to which the propagation effects be responsible in reproducing the observed anomalies. It will also be interesting to examine how non-dipolar components of stellar magnetic fields affect the emission. Such efforts (both observation and simulation) will be crucial in characterizing the phenomenon, and may also help us in understanding the emission produced by cooler objects.

\begin{acknowledgements}
We thank the referee for very useful comments.
We thank budding artist Tanay Singh for creating the illustration showing propagation effects.
BD acknowledges support from the Bartol Research Institute. 
We thank the staff of the GMRT and the National Radio Astronomy Observatory (NRAO) that made our observations possible.
The GMRT is run by the National Centre for Radio Astrophysics of the Tata Institute of Fundamental Research.
The National Radio Astronomy Observatory is a facility of the National Science Foundation operated under cooperative agreement by Associated Universities, Inc.
This research has made use of NASA's Astrophysics Data System.
\end{acknowledgements}

\bibliography{das}
\bibliographystyle{aasjournal}

\appendix
\begin{figure*}
    \centering
    \includegraphics[width=0.8\textwidth]{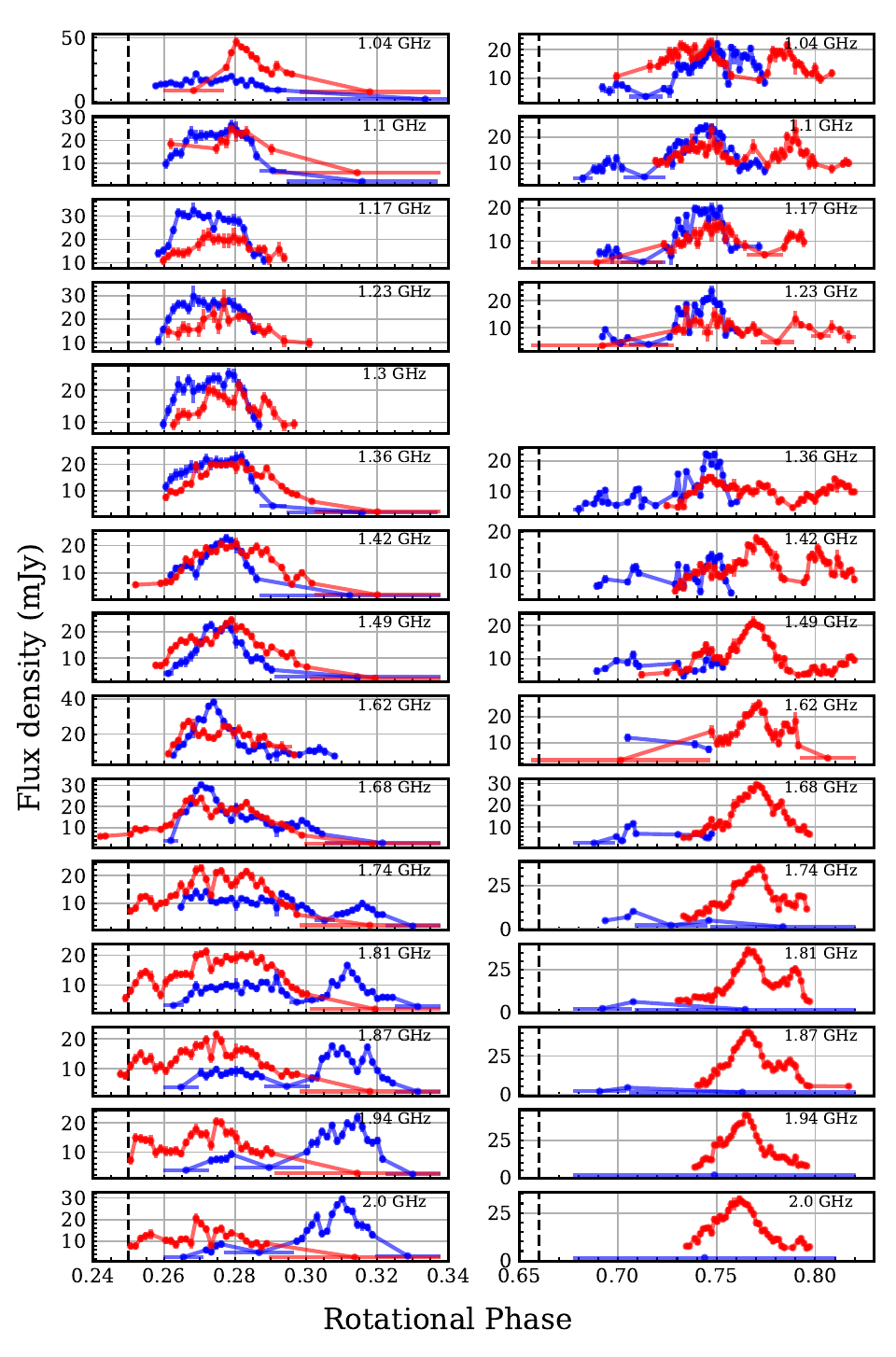}
    \caption{The lightcurves over 1--2 GHz (VLA L band). Red and blue represent RCP and LCP respectively. The missing panel in the right column is due to the fact that the corresponding data had to be completely flagged due to radio frequency interference (RFI). The dashed vertical lines mark null 1 (left panels) and null 2 (right panels). For definitions of null 1 and null 2, see \S\ref{sec:observation}.}
    \label{fig:hd142990_L_full_lc}
\end{figure*}

\begin{figure*}
    \centering
    \includegraphics[width=0.8\textwidth]{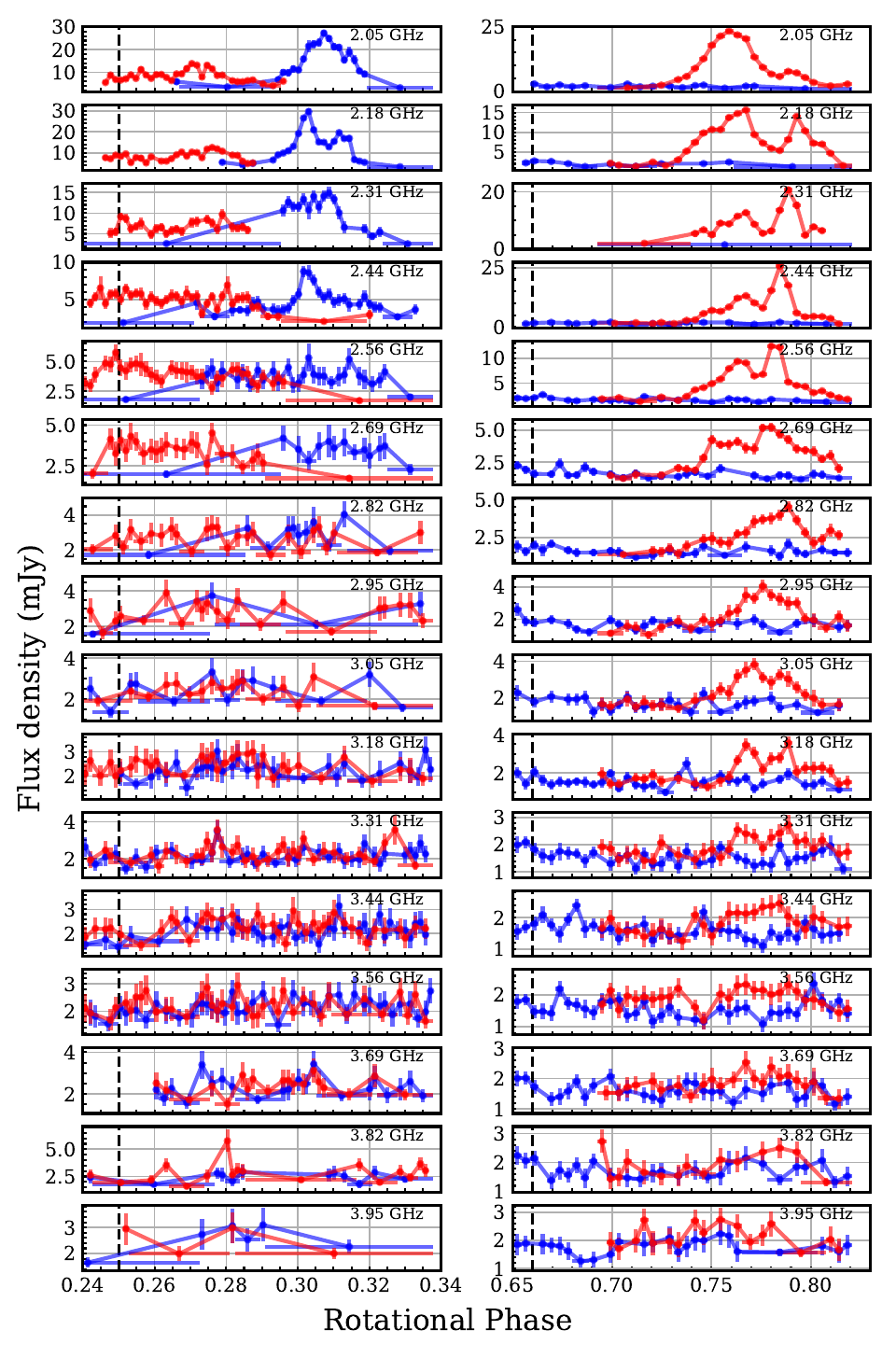}
    \caption{The lightcurves over 2--4 GHz (VLA S band). Red and blue represent RCP and LCP respectively. The dashed vertical lines mark null 1 (left panels) and null 2 (right panels). For definitions of null 1 and null 2, see \S\ref{sec:observation}.}
    \label{fig:hd142990_S_full_lc}
\end{figure*}


\end{document}